\documentclass[aps, reprint, superscriptaddress, nofootinbib]{revtex4-1}
\usepackage{amsmath, latexsym, amssymb, hyperref, graphicx, color, slashed}
\definecolor{nicered}{rgb}{0.7,0.1,0.1}
\definecolor{nicegreen}{rgb}{0.1,0.5,0.1}
\hypersetup{colorlinks, citecolor=nicegreen,linkcolor=nicered}

\begin{document}

\newcommand{\Sec}[1]{ \medskip \noindent {\sl \bfseries #1}}
\newcommand{\subsec}[1]{ \medskip \noindent {\sl \bfseries #1}}
\newcommand{\Par}[1]{ \medskip \noindent {\em #1}}

\title{Right-handed lepton mixings at the LHC}
\author{Juan Carlos \surname{Vasquez}}
\affiliation{SISSA/INFN,  via Bonomea, 265 - 34136 Trieste, Italy.}
\affiliation{  ICTP, Strada Costiera, 11I - 34151 Trieste, Italy.}
\affiliation{Gran Sasso Science Institute, Viale Crispi 7, 67100 LAquila, Italy. }
\date{\today}

\begin{abstract}
   We study how the elements of the  leptonic right-handed mixing matrix can be determined at the LHC in the minimal Left-Right symmetric extension of the standard model. We do it by explicitly relating them with physical quantities of the Keung-Senjanovi\'c  process and the lepton number violating decays of the right doubly charged scalar.  We also point out that the left and right doubly charged scalars can be distinguished at the LHC,  without measuring the polarization of the final state leptons coming from their decays. 
\end{abstract}

\maketitle

%
\section{Introduction}

The Left-Right  symmetric theory is based on the gauge group  $SU(2)_L \times 
SU(2)_R \times U(1)_{B-L}$ \cite{lrmodel,GoranNuclphys79}, times a Left-Right symmetry that may be generalized parity ($\mathcal{P}$) or charge conjugation ($\mathcal{C}$)  (for reviews see \cite{Senjanovic:2011zz}). It introduces  three new heavy
gauge bosons
$W^{+}_R$, $W^{-}_R$, $Z_R$ and the heavy  neutrino states $N$.  
In this model,  the maximally observed
parity non 
conservation is a low energy phenomenon, which ought to disappear at energies
above
the $W_R$ mass. Furthermore,  the smallness of neutrino masses is related to the near maximality of parity violation \cite{Minkowski:1977sc,Mohapatra:1979ia, Mohapatra:1980yp}, through the seesaw mechanism \cite{Minkowski:1977sc,Mohapatra:1979ia, Mohapatra:1980yp,see-sawothers1}.

Theoretical  bounds on the Left-Right scale were considered in the past. The small $K_L-K_R$ mass difference gives a lower bound on the Left-Right-scale of around 3 TeV in the minimal model \cite{Beall:1981ze}. More recently in \cite{Bertolini:2014sua}, an updated study and a complete gauge invariant computation of the $K_L,K_S$ and $B_d,B_s$ meson parameters,  gives  $m_{W_R}> 3.1(2.9)$ TeV for $\mathcal{P}(\mathcal{C})$. In \cite{Maiezza:2014ala} it is claimed that for parity as the Left-Right symmetry,  the $\theta_{QCD}$ parameter, together with K-meson mass difference $\Delta M_K$,  push the mass of $W_R$ up to 20 TeV \cite{Bertolini:2014sua,Maiezza:2014ala};  however this depends on the UV completion of the theory.  Direct  LHC 
searches, on the other hand, gives in
some channels
a lower bound of around $ 3$ TeV \cite{Khachatryan:2014dka}.

It turns out that
there exists \cite{Keung:1983uu} 
an exiting decay  of $W_R$  into two charged leptons and
two jets ($ W_R \rightarrow l + N \rightarrow ll + jj$). We  refer to it as the Keung-Senjanovi\'c (KS) process. This process has a small background 
and  no missing energy. 
It  gives a clean signal for the $W_R$ production 
at LHC, as well as
probing  the Majorana masses of 
the heavy  neutrinos.
Since there is no missing energy in the decay, the reconstruction of
the
 $W_R$ and $N$ invariant masses is possible. If true, the Majorana mass of
$N$ will lead to 
the decay of the heavy  neutrino into a charged lepton and  two jets 
($ N \rightarrow l + jj$),  with the same probability of decaying into
a lepton or antilepton. Recetly CMS gives and excess in the ee-channel of 2.8$\sigma$ for  this particular process at $m_{eejj}\approx2.1$TeV   \cite{Khachatryan:2014dka}.  Several works  have been proposed \cite{Aguilar-Saavedra:2014ola,Aguilar-Saavedra:2014ola1,Heikinheimo:2014tba,Gluza:2015goa,Dobrescu:2015qna,Coloma:2015una,Bandyopadhyay:2015fka,Dev:2015pga,Brehmer:2015cia} in order to explain this excess and  the conclusion was that it  would need a higher Left-Right symmetry breaking scale, or  a more  general  mixing scenario with pseudo-Dirac heavy neutrinos. Next LHC run will be crucial to establish or discard this excess.

The production of $W_R$ is ensured at the LHC because in the quark sector the left and right mixing matrices are related. For $\mathcal{C}$ as the Left-Right symmetry, the mixing angles are exactly equal, therefore the production rate of $W_R$ is the same as the one of $W$. For $\mathcal{P}$  the situation is more subtle and needed an in-depth study. Finally in \cite{Senjanovic:2014pva}  a simple analytic expression valid in the entire parameter space 
was derived for the right-handed quark mixing matrix. It turns out that despite parity being maximally broken in nature, the Right and Left quark mixing matrices end up being very similar. Moreover the hypothesis of equal mixing angles  can be tested  at the LHC by studying the hadronic decays of $W_R$ \cite{Fowlie:2014mza}.

In the Leptonic sector the connection between the Left and Right leptonic mixing matrices goes away,  since light and heavy neutrino masses are different.  For $\mathcal{C}$ as the Left-Right symmetry, the  Dirac masses of neutrinos are unambiguously  determined in terms of the heavy and light neutrino masses \cite{Nemevsek:2012iq}. Light neutrino masses are probed by low energy experiments, whereas the ones of the heavy neutrinos can be determined at the LHC.  This is why  the precise determination of the right-handed leptonic mixing matrix, the main topic of this work, is of fundamental importance.

 We focus on the determination of the elements of the leptonic mixing matrix $V_R$ at the LHC, through the  KS process and the decays of the doubly-charged scalar $\delta^{++}_{R}$  belonging to the $SU(2)_R$ triplet. We point out that these two processes are not sensitive to three of the  phases appearing in $V_R$, unlike electric dipole moments of charged leptons. 

The rest of this paper is organized as follows.  In Section 2 we give a brief description of the model and the main relevant interactions for our purposes. In Section 3  we show the determination of the three mixing angles and the ``Dirac" type phase appearing in $V_R$. We do it in terms of physical observables in the KS process. We also show for $\mathcal{C}$ as the Left-Right symmetry, how the branching ratios of the doubly charged scalar $\delta_R^{++}$ into $e^+e^+$, $e^+\mu^+$ and $\mu^+\mu^+$ can be used to determine the Majorana type phases. We consider for illustration the type II see-saw dominance and  put some representative values for the ``Dirac" phase, the lightest and the heaviest right-handed neutrino masses. Finally,  we also show that the doubly charged scalars $\delta_L^{++}$ and $\delta_R^{++}$ may be distinguished at the LHC, without measuring the polarization of the charged leptons coming from their decays.

%
%

\section{ The minimal Left-Right symmetric model}
The minimal Left-Right symmetric model \cite{lrmodel,GoranNuclphys79} is based on the gauge group $\mathcal{G}=SU(2)_L\times SU(2)_R\times U(1)_{B-L}$, with an additional discrete symmetry that may be generalized parity ($\mathcal{P}$) or  charge conjugation ($\mathcal{C}$).
Quarks and leptons are assigned to be doublets in the following irreducible representations 
of the gauge group:
 \begin{align*}
q_L =
\left( \begin{array}{ccc}
u  \\
d \\
\end{array} \right) _L :   (2,1,\frac{1}{3}), \quad  q_R
=\left( \begin{array}{ccc}
u  \\
d \\
\end{array} \right) _R :   (1,2,\frac{1}{3}),
\end{align*}

\begin{align}
\
L_L =
\left( \begin{array}{ccc}
\nu  \\
l \\
\end{array} \right) _L :   (2,1,-1),\quad  L_R =\left(
\begin{array}{ccc}
N  \\
l \\
\end{array} \right) _R :  (1,2,-1). \nonumber \\
\end{align}

 $N$ represents the new heavy neutrino states, whose presence play a crucial role in explaining the smallness of the neutrino masses on the basis of the see-saw mechanism.

The Higgs sector sector \cite{Minkowski:1977sc,Mohapatra:1979ia} consists in one bidoublet $\Phi$, in the (2,2,0) representation of $\mathcal{G}$,  two scalar triplets $\Delta_L$ and $\Delta_R$, belonging to (3,1,2) and (1,3,2) representation respectively
 \begin{align}
\Phi =
\left( \begin{array}{ccc}
\phi_1^0 && \phi_2^+  \\
\phi_1^- && \phi_2^0 \\
\end{array} \right) , \quad  \Delta_{L,R}=\left( \begin{array}{ccc}
\delta^+_{L,R}/\sqrt{2} && \delta_{L,R}^{++}  \\
\delta_{L,R}^{0} &&- \delta^+_{L,R}/\sqrt{2}  \\
\end{array} \right). \nonumber \\
\end{align}

 Under the discrete left-right symmetry the fields transform as follows:
 \begin{widetext}
\begin{align}
\mathcal{P} :  \left\{ \begin{array}{ll}
 \mathcal{P}f_L\mathcal{P}^{-1} =  \gamma_0f_R  \\
 \mathcal{P}\Phi\mathcal{P}^{-1} =\Phi^{\dagger} \\
 \mathcal{P}\Delta_{(L,R)}\mathcal{P}^{-1} = -\Delta_{(R,L)}  
\qquad
 \end{array} \right. \quad \mathcal{C} :  \left\{ \begin{array}{ll}
 \mathcal{C}f_L  \mathcal{C}^{-1} =C(\bar{f_R})^T  \\
 \mathcal{C}\Phi\mathcal{C}^{-1} = \Phi^{T} \\
 \mathcal{C}\Delta_{(L,R)}\mathcal{C}^{-1} = -\Delta^*_{(R,L)} 
 \end{array} \right.  \label{relations}
 \end{align}
 \end{widetext}
 where $\gamma_{\mu}$ ($\mu=0,1,2,3.$) are the gamma matrices and $C$ is the charge conjugation operator.

Lepton masses are due to the following Yukawa interactions (once the Higgs fields take their v.e.v along their neutral components)
\begin{align}
\mathcal{L}_Y=&\bar{L}_L(Y_{\Phi}\Phi+\tilde{Y}_{\Phi}\tilde{\Phi})L_R +\frac{1}{2}(L_L^TCi\sigma_2Y_{\Delta_L}\Delta_LL_L \nonumber \\ 
&+L_R^TCi\sigma_2Y_{\Delta_R}\Delta_RL_R )+h.c., \label{yukawas}
\end{align}
where $\tilde{\Phi}=\sigma_2\Phi^*\sigma_2$, $\sigma_2$ being the Pauli matrix.

 Invariance of the Lagrangian under the Left-Right symmetry requires
 \begin{align}
 \mathcal{P} :  \left\{ \begin{array}{ll}
 Y_{\Delta_{R,L} } = Y_{\Delta_{L,R}} \\
 Y_{\Phi}=Y_{\Phi}^{\dagger} \\
 \tilde{Y}_{\Phi}=\tilde{Y}_{\Phi}^{\dagger}
 \end{array} \right.,\quad
\mathcal{C} :  \left\{ \begin{array}{ll}
 Y_{\Delta_{R,L} } = Y^*_{\Delta_{L,R}} \\
 Y_{\Phi}=Y_{\Phi}^T \\
 \tilde{Y}_{\Phi}=\tilde{Y}_{\Phi}^T
 \end{array} \right.
 \end{align}
 

The v.e.v's of the Higgs fields may be written as \cite{Mohapatra:1980yp}
 \begin{align*}
\langle \Phi \rangle =
\left( \begin{array}{ccc}
v_1 && 0  \\
0 && v_2 e^{i\alpha} \\
\end{array} \right).
\end{align*}

 \begin{align}
\langle \Delta_{R} \rangle=\left( \begin{array}{ccc}
0&&0 \\
v_R  &&0  \\
\end{array} \right)  , \quad  \langle \Delta_{L} \rangle=\left( \begin{array}{ccc}
0&&0 \\
v_Le^{i\theta_L} &&0  \\
\end{array} \right)
\end{align}
 where  $v_L\propto (v_1^2+v_2^2)/v_R$ and the neutrino masses take the see-saw form \cite{Mohapatra:1979ia}
\begin{align}
& M_{N}=Y_{\Delta_R}^*v_R, \\
&M_{\nu}=Y_{\Delta_L}v_Le^{i\theta_L}-M^{\dagger}_D\frac{1}{M_N}M_D^*, \\
&M_D=v_1Y_{\Phi}+\tilde{Y}_{\Phi}v_2 e^{-i\alpha} 
\end{align}
The charged lepton mass matrix is 
\begin{align}
&M_l= Y_{\Phi}v_2e^{i\alpha}+\tilde{Y}_{\Phi}v_1
\end{align}
$\alpha$ is called the ``spontaneous'' CP  phase.  All  the physical effects due to $\theta_L,$  can be neglected, since this phase is always accompanied by the small $v_L$.

As usual, the mass matrices can be diagonalized  by the biunitary transformations
\begin{align}
&M_l = U_{lL}m_lU^{\dagger}_{lR}, \quad M_{D}=U_{D L}m_{D}U^{\dagger}_{DR}, \nonumber \\ 
 &M_{\nu}=U^{*}_{\nu }m_{\nu}U^{\dagger}_{\nu},\quad M_{N}=U^{*}_{N }m_{N}U^{\dagger}_{N},
\end{align}
where $m_l$, $m_{\nu}$ and $m_{N}$ are diagonal matrices with real, positive eigenvalues.

In the mass eigenstate basis the flavor changing charged current Lagrangian is
\begin{equation}
\mathcal{L}_{cc}= \frac{g}{\sqrt{2}}(\bar{\nu}_L V_{L}^\dag \slashed{W}_{\!L} l_L+\bar{N}_R V_{R}^\dag \slashed{W}_{\!R} l_R) +h.c.,\label{cclagrangian}‏
\end{equation}
 $V_L$ and $V_R$ are the left and right leptonic mixing matrices respectively
\begin{align}                                                                 
& V_L = U^{\dagger}_{lL}U_{\nu}, \\
& V_R = U^{\dagger}_{lR}U^*_{N}.
\end{align}

We may use the freedom of rephasing the charge lepton fields to remove  three unphysical phases from $V_L$, which ends up having 3 mixing angles and 3 phases. As it is well known, the mixing angles of this matrix are probed by low energy experiments. Instead we  focus  in the precise determination of the mixing angles and phases of its right-handed analog ($V_R$) at hadron colliders. This matrix $V_R$ has in general 3 different angles and 6 phases. We  write it in the form $V_R=K_e\hat{V}_RK_N$ where $K_e=\text{diag}(e^{i\phi_e},e^{i\phi_{\mu}},e^{i\phi_{\tau}})$, $K_N=\text{diag}(1,e^{i\phi_{2}},e^{i\phi_{3}})$  and

\begin{widetext}
\begin{align} 
\hat{V}_R=\left( \begin{array}{ccccc}
c_{13}c_{12}& c_{13}s_{12} & s_{13}\\
-s_{12}c_{23}e^{i\delta}-c_{12}s_{13}s_{23} & c_{12}c_{23}e^{i\delta}-s_{12}s_{13}s_{23} & c_{13}s_{23}  \\
s_{12}s_{23}e^{i\delta}-c_{12}s_{13}c_{23}& -c_{12}s_{23}e^{i\delta}-s_{12}s_{13}c_{23}& c_{13}c_{23}  \\
\end{array} \right),\nonumber \\ \label{PMNS}
\end{align}
\end{widetext}
 $s_{\alpha\beta}$ is the short-hand notation for $\sin\theta_{\alpha\beta}$ ($\alpha,\beta=1,2,3$).

The next relevant interactions for our discussion are the ones between the charged leptons and the doubly charged scalars
\begin{align}
 &\mathcal{L}_{\Delta} = \frac{1}{2} l_R^TC Y'_{\Delta_R} \delta_R^{++}l_R +
\frac{1}{2} l_L^TC Y'_{\Delta_L} \delta_L^{++}l_L +h.c., \label{Ldelta} \\
 & Y'_{\Delta_R} = \frac{g}{m_{W_R}}V_R^*m_NV_R^{\dagger}. \label{Y_R}
\end{align}

If  $\mathcal{C}$ is the left-right symmetry,  is easy to see from Eqs.  \eqref{relations} and \eqref{yukawas} that \cite{Senjanovic:2011zz}
\begin{equation}
 Y'_{\Delta_L} = (Y'_{\Delta_R})^*. \label{relacionCdeltas}
\end{equation}

 For parity ($\mathcal{P}$) the situation is different since for a non-zero   spontaneous phase  the charged lepton masses are not hermitian. Then after the symmetry breaking, one would expect  that the left and right Yukawa interactions  with the doubly-charged scalar are not the same.  It turns out that for right-handed neutrinos masses accessible at the LHC, the charged lepton mass matrices end up being quite hermitian \cite{trabajoGVN}. Let us notice that it implies that Yukawas of the doubly charge scalars must satisfy
 \begin{align}
 &Y^{'}_{\Delta_L}= S_lY^{'}_{\Delta_R} S_l+i\tan \beta \sin\alpha(R^*Y'_{\Delta_R}S_l+S_lY'_{\Delta_R}R^{\dagger})\nonumber \\ &+\mathcal{O}\left[(\tan\beta\sin\alpha)^2\right] \label{relacionPdeltas}
 \end{align} 
 with
 \begin{equation}
  (R)_{ij}= \frac{(M'_D)_{ij}}{(m_l)_i+(m_l)_j}-\frac{1}{2}\tan\beta e^{-i\alpha}(S_l)_{ij}
 \end{equation}

 where $S_l$ is a $3\times3$ matrix with $\pm$ signs in the diagonal entries and zero otherwise, $M'_D=U_{lL}^{\dagger}M_DU_{lR}$  and  $\beta\equiv v_2/v_1$. This is obtained in analogy to the approach used  for the quark mixing matrix in \cite{Senjanovic:2014pva,Senjanovic:2015yea}, where  it is also shown that $\tan 2\beta\sin\alpha \lesssim 2m_b/m_t$. Therefore one can safely assume that  $Y^{'}_{\Delta_L}\backsimeq Y^{'}_{\Delta_R}$ as a leading order approximation in the most interesting scenario.

Notice that \eqref{Y_R} depends on the Majorana phases. Therefore the decay rates of $\delta_R^{++}$ into two leptons in the final state depend in a CP-even way on the Dirac and Majorana phases.

%
%
\section{Determination of the right-handed leptonic mixing matrix }
In this section we show how the three angles $\theta_{12}$,$\theta_{23}$,$\theta_{13}$ and the Dirac phase $\delta$, appearing in $V_R$  are all expressed in term of physical observables at the LHC. More precisely,  we find  analytic expressions relating the elements of $\hat{V}_R$ with some physical branching ratios of the KS process. For the Majorana phases we point out that they can be obtained through the decays of the doubly charged scalar. Moreover these measurements could serve as a cross-checking for the model.

\subsection{Keung-Senjanovi\'c process}
We begin our analysis by considering the KS
process. It has a clean signal with almost no background  that consists
in two leptons and two jets in the final state. This process has the striking features 
of  no missing energy in  the final state and the  amplification by the $W_R$ resonance.  Measuring the energy and momenta of the final particles  it allows the full reconstruction of the masses of the $W_R$ and the heavy neutrino $N$. Studies of this process were performed in the past \cite{Ferrari:2000sp},   with the conclusion that $W_R$ can be discovered at the LHC with a mass up to $\simeq 6$ TeV, masses for the right-handed neutrinos of the order $m_N\simeq100 \text{GeV- } 1 \text{TeV}$  for 300 $\text{fb}^{-1}$ of integrated luminosity. In  \cite{Gopalakrishna:2010xm,Han:2012vk}  completed studies of the $W_R$ production and decays at the LHC were done. They gave special emphasis to the chiral couplings of the $W_R$ with initial and final state quarks as well as the final state leptons.  They showed that it is possible to determine (by studying angular correlations and asymmetries between the participating particles) the chiral properties of $W_R$ and the fermions.

\begin{figure}
\includegraphics[width=0.42\textwidth]{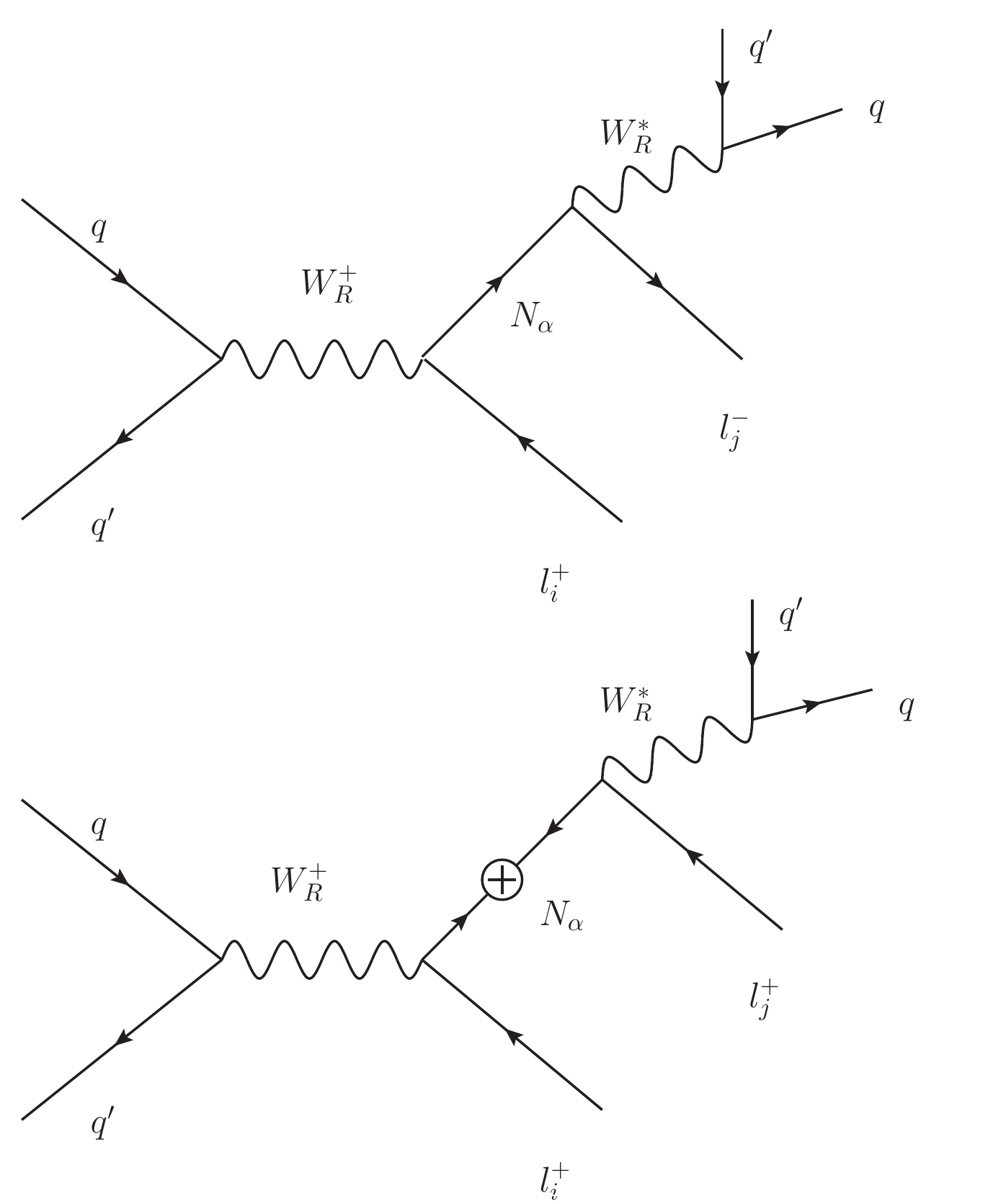}  
\caption{\label{fig1}   Keung-Senjanovi\'c  process in both opposite-sign leptons (Top) and the lepton-number-violating
same-sign leptons in the final state (Bottom).}
\end{figure}

The KS process offers also the possibility of observing  both the restoration of the Left-Right symmetry and the
Majorana nature of neutrinos at colliders (see FIG. \ref{fig1}). The latter implies the equality between the decay rates in the  same-sign and the  opposite-sign leptons in the final state.

Once $W_R$ is produced on-shell,  it  decays into a lepton and the heavy neutrino
$N$. If the  $W_R$ mass is bigger than the masses of the $N_{\alpha}$ for  $\alpha =
1,2,3$,  the decay rate of $W_R \rightarrow l_il_k jj$ is (no summation over repeated indices)
\begin{align}
&\Gamma(W_R^+ \rightarrow l_i^+ l_k^+ jj)=\sum_{q q'}\Gamma(W_R^+ \rightarrow l_i^+ l_k^+ qq^{'})          \nonumber \\
=&\sum_{q q'}\Gamma(W_R^+ \rightarrow l_i^+
N_{\alpha})\text{Br}(N_{\alpha}\rightarrow l_k^+qq^{'}), \nonumber \\ \label{factorization}
\end{align}
where $i,k= e,\mu,\tau$ .


Notice that if the heavy neutrino masses are not degenerate,  in
general the KS process is sensitive only to the Dirac type phase $\delta$.
In this case both lepton number conserving and lepton number violating
channels  give  the same results. The partonic  processes are illustrated in FIG. \ref{fig1}. 

For degenerate heavy neutrino masses, one may easily see from the same-sign
leptons in the final state,  that there is a CP-even dependence on the 
phases in $K_N$. Notice that this channel breaks the total
lepton number,  then is clear that we should have some dependence on the
Majorana phases.   In the case of at least two degenerate heavy neutrino masses, it is in principle possible to construct CP-odd, triple-vector-product asymmetries with three momenta or any mixture of momenta and spin for the participating particles.

From Eq. \eqref{cclagrangian} we find that the decay rate of $W_R^+ \rightarrow l_i^+ N_{\alpha}$ is
\begin{equation}
 \Gamma(W_R^+ \rightarrow l_i^+ N_{\alpha})=\frac{g^2}{8\pi}|(V_R^{\dagger})_{\alpha i}|^2\frac{|\vec{p}^{\alpha}_2|^2}{m_{W_R}^2}[\frac{|\vec{p}^{\alpha}_2|}{3}+E_2^{\alpha}], \label{N}
 \end{equation}
 $\vec{p}^{\alpha}_2$  is the momentum of the right-handed neutrino $N_{\alpha}$. $E_2^{\alpha}$ is the energy of $N_{\alpha}$ and $\vec{p}^{\alpha}_2$ is such that
 \begin{equation}
 |\vec{p}^{\alpha}_2|+\sqrt{|\vec{p}^{\alpha}_2|^2+m_{N_{\alpha}}^2}=m_{W_R}.
 \end{equation}
 
  The 3-body decay of $N$ into one lepton and two jets is given by
\begin{align}
\Gamma(N_{\alpha} \rightarrow l_k^+jj)&= \nonumber \\ &2N_c(\frac{g^2}{8m_{W_R}^2})^2\frac{m^5_{N_{\alpha}}}{192\pi^2}|(V_R^{\dagger})_{\alpha k}|^2 \sum_{q q'}|(V^Q_R)^{\dagger}_{qq'}|^2, \nonumber \\ \label{3N}
\end{align}
where $V_R^{Q}$ is the right-handed quark mixing matrix,  $N_C$ is the number of colors and the sum over $q,q'$ includes the kinematically allowed heavy neutrino decays.

For heavy neutrinos masses above the pion threshold, the dominant decay rate are the hadronic ones and in this case  the following ratio takes the simple form 
\begin{widetext}
\begin{align}
\frac{\Gamma(W_R^+ \rightarrow N_{\alpha}l_i \rightarrow l_i^+ l_k^+ jj)}{\Gamma(W_R^+\rightarrow  N_{\alpha'}l_r \rightarrow l_r^+ l_s^+ jj)}=\frac{\sigma(pp\rightarrow W_R^+ \rightarrow N_{\alpha}l_i \rightarrow l_i^+ l_k^+ jj)}{\sigma(pp\rightarrow W_R^+\rightarrow  N_{\alpha'}l_r \rightarrow l_r^+ l_s^+ jj)} =\frac{|(V_R^{\dagger})_{\alpha i}|^2|(V_R^{\dagger})_{\alpha k}|^2c^{\alpha}}{|(V_R^{\dagger})_{\alpha'r}|^2|(V_R^{\dagger})_{\alpha's}|^2c^{\alpha'}}, \label{12}
\end{align}
\end{widetext}
where
\begin{figure*}
\includegraphics[width=6in]{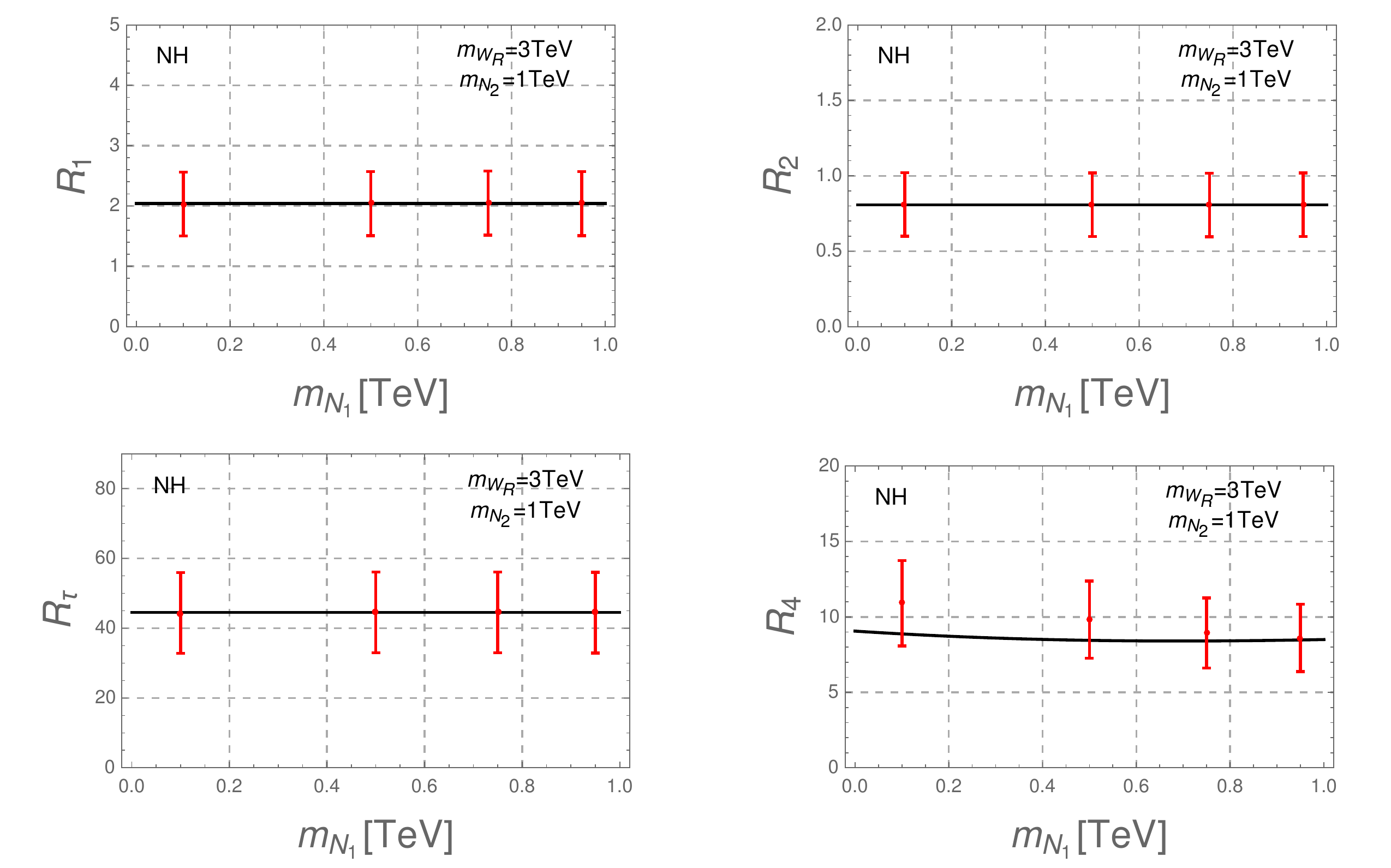}
\caption{ \label{fig2N}\small
Plots for the quantities $R_1$,$R_2$,$R_{\tau}$ and $R_4$ as a function of the lightest neutrino mass eigenstate in the NH case. Red dots with errors bars are the results obtained by taking into account the hadronization effects using Pythia 6. We assume the values of the gauge boson $m_{W_R}=3$ TeV and the heavy neutrino mass  $m_{N_2}=1$ TeV  }
\end{figure*}
\begin{equation}
 c^{\alpha} \equiv  |\vec{p}^{\alpha}_2|^2[\frac{|\vec{p}^{\alpha}_2|}{3}+E_2^{\alpha}],
\end{equation}
 all the hadronic and quark mixing part cancels and we end up having a quantity that depends only on the physical masses and  the elements of $V_R$.   When $\alpha=\alpha'$ the expression further simplifies and depends only on the elements of $V_R$. 

 In what follows we consider the case when one, two or three heavy neutrinos  are accessible at the LHC.
 \begin{figure*}
\includegraphics[width=6in]{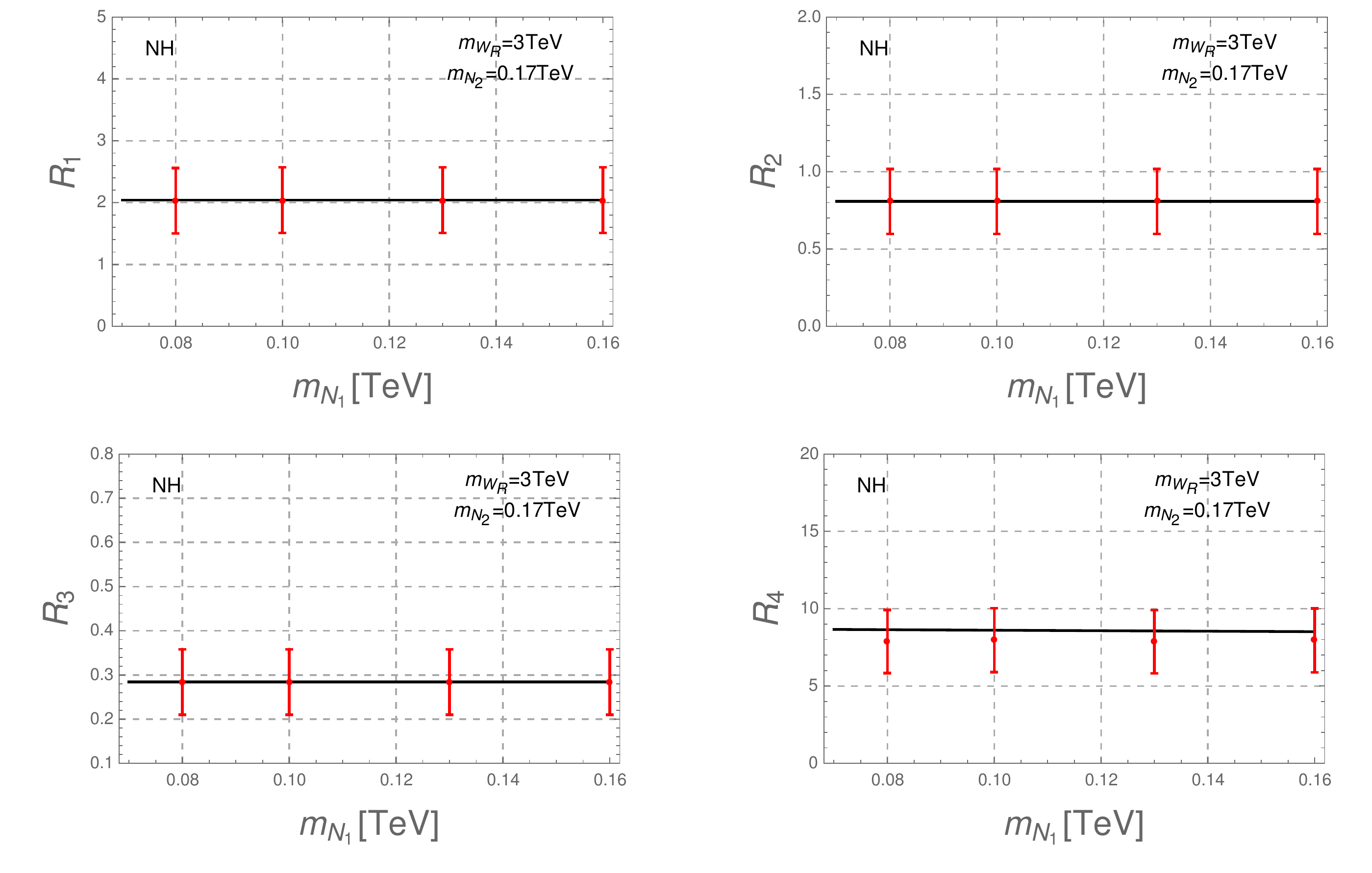}
\caption{ \label{fig3N_NH}\small
Plots for the quantities $R_1$,$R_2$,$R_{3}$ and $R_4$ as a function of the lightest neutrino mass eigenstate in the NH case. Red dots with errors bars are the results obtained by taking into account the hadronization effects using Pythia 6. We assume the values of the gauge boson $m_{W_R}=3$ TeV and the heavy neutrino mass  $m_{N_2}=0.17$ TeV  }
\end{figure*}
\begin{figure*}
\includegraphics[width=6in]{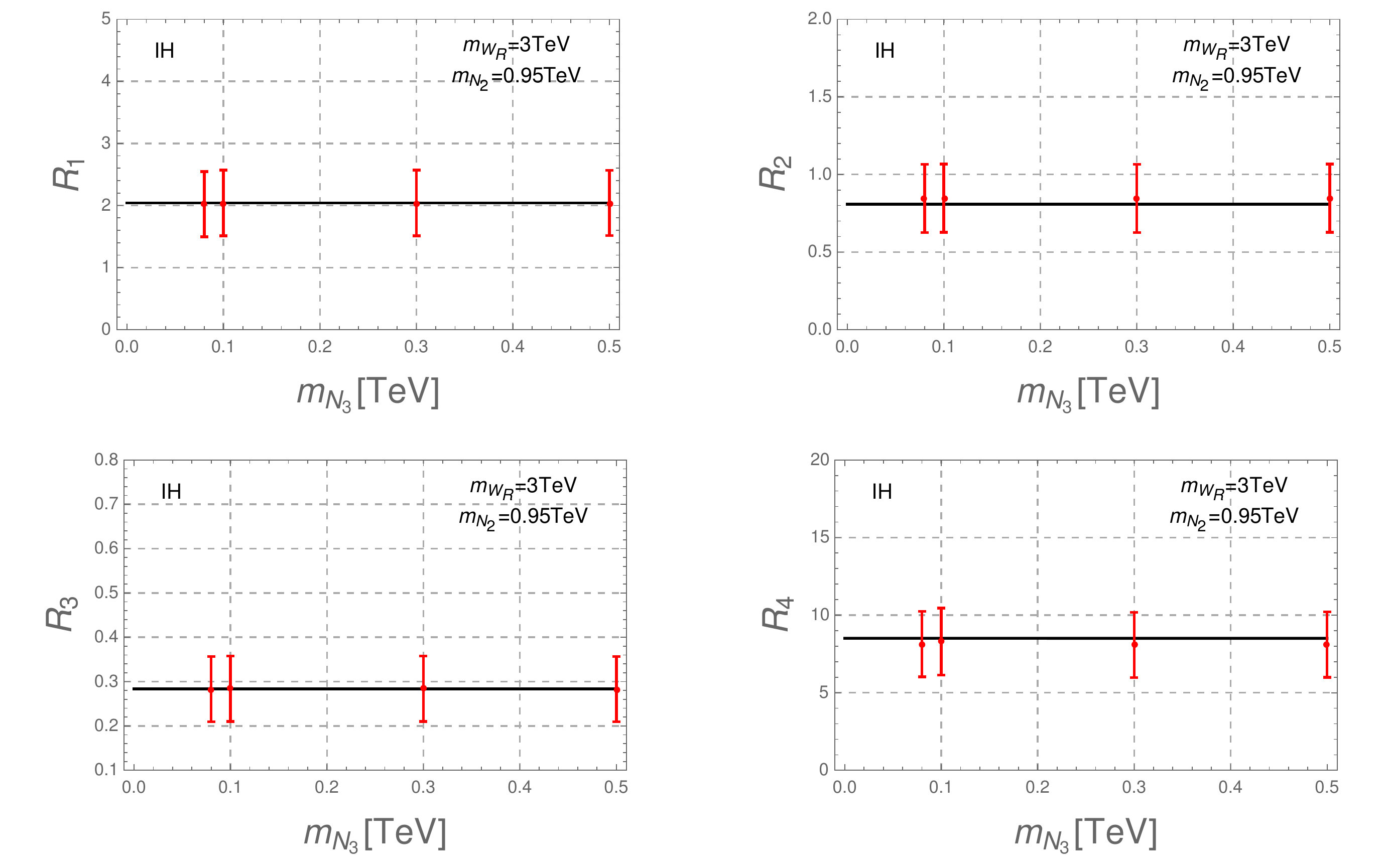}
\caption{ \label{fig3N_IH}\small
Plots for the quantities $R_1$,$R_2$,$R_{3}$ and $R_4$ as a function of the lightest neutrino mass eigenstate in the IH case. Red dots with errors bars are the results obtained by taking into account the hadronization effects using Pythia 6. We assume the values of the gauge boson $m_{W_R}=3$ TeV and the heavy neutrino mass  $m_{N_2}=0.95$ TeV  }
\end{figure*}

 \textbf{One heavy neutrino case}:  it may happen that even if the $W_R$ is found at the LHC,  just one of the heavy neutrino mass can be reconstructed. In this case  we  see from Eq. \eqref{12} (taking $r=s=\mu$) that there are only two independent quantities including tau leptons in the final state.   
 
 If only electrons and muons are considered is easy to see that there is only one independent quantity within this analysis.

 \textbf{Two heavy neutrinos case}: one expect for  two heavy neutrino at the LHC, that in order to probe all the elements of the mixing matrix $V_R$ the decays  of the heavy neutrinos $N$ into electrons, muons and tau leptons must be identified. In fact, in this case analytical solutions for the three mixing angles and the Dirac phase $\delta$ can be found in terms of physical quantities at the LHC,  this can be seen  by considering  $\alpha=\alpha'$ in Eq. \eqref{12}, namely
\begin{align}
&\frac{\Gamma(W_R^+ \rightarrow N_{\alpha}e^+ \rightarrow e^+\mu^+ jj)}{\Gamma(W_R^+\rightarrow  N_{\alpha}\mu^+\rightarrow \mu^+ \mu^+ jj)}=\frac{|(V_R^{\dagger})_{\alpha e}|^2}{|(V_R^{\dagger})_{\alpha \mu}|^2}\equiv R_{\alpha} \label{14}\\
\text{where} \nonumber  \\ 
&\alpha=1,2. \nonumber
\end{align}

There are 4 unknown parameters in $\hat{V}_R$ ($\theta_{12}$, $\theta_{13}$, $\theta_{23}$ and $\delta$). By using the above ratios it is possible to probe  2 of them.  There is just another independent quantity considering electron and muons in the final state
 \begin{align}
&\frac{\Gamma(W_R^+ \rightarrow N_1e^+ \rightarrow e^+e^+ jj)}{\Gamma(W_R^+\rightarrow  N_2e^+\rightarrow e^+ e^+ jj)} \equiv R_4= \frac{|(V_R^{\dagger})_{1e}|^4c^{(1)}}{|(V_R^{\dagger})_{2e}|^4c^{(2)}}. \label{2family}
\end{align}
So we conclude that in order to probe the three mixings angles and the Dirac phase with 2 heavy neutrinos on-shell, tau leptons must be included into the  analysis and to this end consider the following relation
\begin{equation}
 \frac{\Gamma(W_R^+\rightarrow N_1 e^+ \rightarrow e^+ e^+ jj)}{\Gamma(W_R^+\rightarrow N_1 e^+ \rightarrow e^+ \tau^+ jj )}=\frac{|(V^{\dagger}_R)_{1e}|^2}{|(V^{\dagger}_R)_{1\tau}|^2}\equiv R_{\tau} \label{theta12}
\end{equation}
and the mixings angles are given by

\begin{align}
& s_{12}^2= \frac{1}{\sqrt{\frac{c^{(2)}}{c^{(1)}}R_4}+1}, \quad s_{13}^2=\frac{-\frac{R_{\tau} R_1}{\sqrt{\frac{c^{(2)}}{c^{(1)}}R_4}}+R_1+R_{\tau}}{R_{\tau} R_1+R_1+R_{\tau}}, \nonumber \\
& \qquad \quad s_{23}^2=\frac{\left(\frac{1}{R_{\tau}}+\frac{1}{R_2}+1\right) \sqrt{\frac{c^{(2)}}{c^{(1)}}R_4}}{\sqrt{\frac{c^{(2)}}{c^{(1)}}R_4}+1}-\frac{1}{R_2}.\label{mixings2HN}
\end{align}
Perhaps the more important  advantage of the above expressions is that they allow a simple interpretation of the three leptonic mixing angles in terms of the final states in the KS process. For instance, from \eqref{mixings2HN} we may see that $\theta_{12}$ is maximal when $R_4>>1$ and minimal when $R_4<<1$.  For  $\theta_{13}$  we notice that its value is  maximal whenever $R_1<<1$ or $R_{\tau}<<1$.  Instead it is minimal when the relation $R_1+R_{\tau}= R_1R_{\tau}/\sqrt{\frac{c^{(2)}}{c^{(1)}}R_4}$ is satisfied. Finally $\theta_{23}$ takes its maximal value when $R_4>>1$ and $R_{\tau}>>1$ and its minimal value when $R_4<<1$ and $R_2>>1$.

For the sake of simplicity we show the expression for the Dirac phase $\delta$ in terms of $R_1$ and  the mixing angles and it is given by
\begin{equation}
 \cos\delta = \frac{c_{13}^2c_{12}^2-R_1(c_{23}^2s_{12}^2+c_{12}^2s_{13}^2s_{23}^2)}{2c_{12}c_{23}s_{12}s_{13}s_{23}R_1} \label{cosd}
\end{equation}

In order to see how the above  results are affected once hadronization effects are taken into account, we extent the Feynrules implementation of the mLRSM  in \cite{Roitgrund:2014zka} to include leptonic mixing in the type II see-saw dominance for $\mathcal{C}$ as the LR symmetry, where it can be shown that $V_R=K_e V_L^*$. The events at the parton level are simulated with Madgraph 5 \cite{Alwall:2014hca} and hadronization effects with Pythia 6 \cite{Sjostrand:2006za}. We use the same cuts applied in \cite{Ferrari:2000sp}, namely both jets must have transverse energy grater than 100 GeV and the invariant mass of the two final  leptons  grater than 200 GeV.  We  take $\theta_{12}=35^o$, $\theta_{23}=45^o$, $\theta_{13}=7^o$ and $\delta=0$ in this illustrative example. 

Furthermore,  there is a proportionality between the two neutrino mass matrices 
\begin{equation}
 \frac{M_N}{\langle \Delta_R \rangle} = \frac{M^*_{\nu}}{\langle \Delta_L \rangle^*},
\end{equation}
which implies \cite{Tello:2010am}
\begin{equation}
 \frac{m^2_{N_2}-m^2_{N_1}}{m^2_{N_3}-m^2_{N_1}}= \frac{m^2_{\nu_2}-m^2_{\nu_1}}{m^2_{\nu_3}-m^2_{\nu_1}} \simeq \pm 0.03,
\end{equation}
where the $\pm$ corresponds to normal/inverted (NH/IH) neutrino mass hierarchy  respectively. Notice that once the Left-Right symmetry is discovered,  this possibility can be verify or falsify by the experiments. We show in Fig. \ref{fig2N} in the case of normal hierarchy neutrino mass spectrum and for heavy neutrino masses accessible at the LHC, the results obtained from the simulation, where it can be readily seen that our suggested strategy for measuring the right handed mixing angles  is feasible at hadron colliders such as the LHC and future ones. Notice that for the IH case,  neutrino mass spectra accessible at the LHC would imply that only one or three neutrino masses can be reconstructed. The largest uncertainties in the production cross sections arises from the uncertainties in the proton PDF's and we assume them to be $26\%$ for $m_{W_R}=3$ TeV as reported in \cite{Khachatryan:2014dka}, although in this paper we consider 13 TeV of center of mass energy, one does not expect this result to change considerably. All this assuming 100$\%$ identification of the tau leptons in the final state. This issue and the expected sensitivity to the leptonic mixing angles will be the subject of future work. 

Finally, from table \ref{tableI} in appendix A we see that the smallest cross sections are the ones of the processes involving tau leptons in the final state. This can readily understood as a consequence of the smallness of the $\theta_{13}$ angle. The results obtained are encouraging, we find that   for heavy neutrino masses near or below the TeV range, a luminosity of $224 $ $\text{fb}^{-1}$ is sufficient to measure all the three mixing angles at the LHC. We determine this value of the luminosity by requiring at least 10 events, since a ratio of the signal over the background  equal to five is reach much faster due to the LNV character of the final states.

\textbf{Three heavy neutrinos case}: once again in this case it is possible to find  analytic expressions for the parameters in $V_R$ in terms of the physical quantities defined in Eq. \eqref{12}. The novelty is that no tau leptons must be identified in the final state, hence rendering this  scenario ideal for the LHC; to this end consider  Eqns. \eqref{14}, \eqref{2family}  and 
\begin{equation}
 \frac{\Gamma(W_R^+ \rightarrow N_{3}e^+ \rightarrow e^+\mu^+ jj)}{\Gamma(W_R^+\rightarrow  N_{3}\mu^+\rightarrow \mu^+ \mu^+ jj)}=\frac{|(V_R^{\dagger})_{3 e}|^2}{|(V_R^{\dagger})_{3 \mu}|^2}\equiv R_{3}.
\end{equation}
A straightforward computation gives
\begin{align}
& s_{12}^2=\frac{1}{1+\sqrt{\frac{c^{(2)}}{c^{(1)}}R_4}}, \quad s_{23}^2=\frac{R-1}{R_3-1}, \nonumber \\ 
 & \qquad \qquad  s_{13}^2= \frac{R-1}{R-\frac{1}{R_3}}. \label{angles}
\end{align}
where
\begin{align}
 R\equiv \frac{1}{\sqrt{\frac{c^{(2)}}{c^{(1)}}R_4}+1}\left[\frac{ \sqrt{\frac{c^{(2)}}{c^{(1)}}R_4}}{R_1}+ \frac{1}{R_2}	\right]
\end{align}

One striking feature of the above expressions is that both $\theta_{13}$ and $\theta_{23}$ are near zero whenever R is close to one, and this in turn implies that $R_1$ is must be close to $R_2$. Furthermore $\theta_{23}$ is nearly maximal when  $R_3\approx R$ and this relation  precisely corresponds to the maximal value $\theta_{13}$ when $R_3\approx R$ but its values are close to one.

As it is clear from the above expressions, the elements of $\hat{V}_R$ have in this parametrization   simple relations in terms of  physical observables at the LHC. The precise form of the Dirac phase $\delta$ is shown in \eqref{cosd}.  Notice that for non-degenerate heavy neutrino masses and within this approach one cannot distinguish  $\delta$ from $-\delta$. In this respect  we notice the CP-odd, triple-vector-product asymmetries in   $\mu \rightarrow e \gamma$ decay and $\mu \rightarrow e$ conversion in Nuclei  \cite{Vasquez:2015una} may resolve this ambiguity and could even  discriminate in the most interesting portion of the parameter's space,  between $\mathcal{C}$ or $\mathcal{P}$ as the Left-Right symmetry.

In Figs. \ref{fig3N_NH} and \ref{fig3N_IH} we show the theoretical values  for the quantities defined above as well as the result obtained using Madgraph 5 and Pythia 6 indicated by the red dots with their respective error bars. We do it for both normal and inverted neutrino mass hierarchies and it is clear from the figures that the hadronic corrections to these quantities are under control and assumed to be  $26\%$.

In this case and from tables \ref{tableII} and \ref{tableIII} in appendix B, we find that for the range of heavy neutrino masses considered i.e. heavy nutrino masses near or bellow the TeV range, the required luminosity necessary for the determination of the three mixing angles is $68$ $\text{fb}^{-1}$ and $45 $ $\text{fb}^{-1}$ for the NH and IH cases respectively. Once again and in analogy with the two heavy neutrinos case,  we find this  value for the luminosity by requiring at least 10 events in the final state, since the  ratio of the signal over the background  equal to five is reach much faster due to the LNV character of the final states. 
\subsection{Decays of the doubly-charged scalar $\delta_R^{++}$}
\begin{figure}
\includegraphics[width=0.42\textwidth]{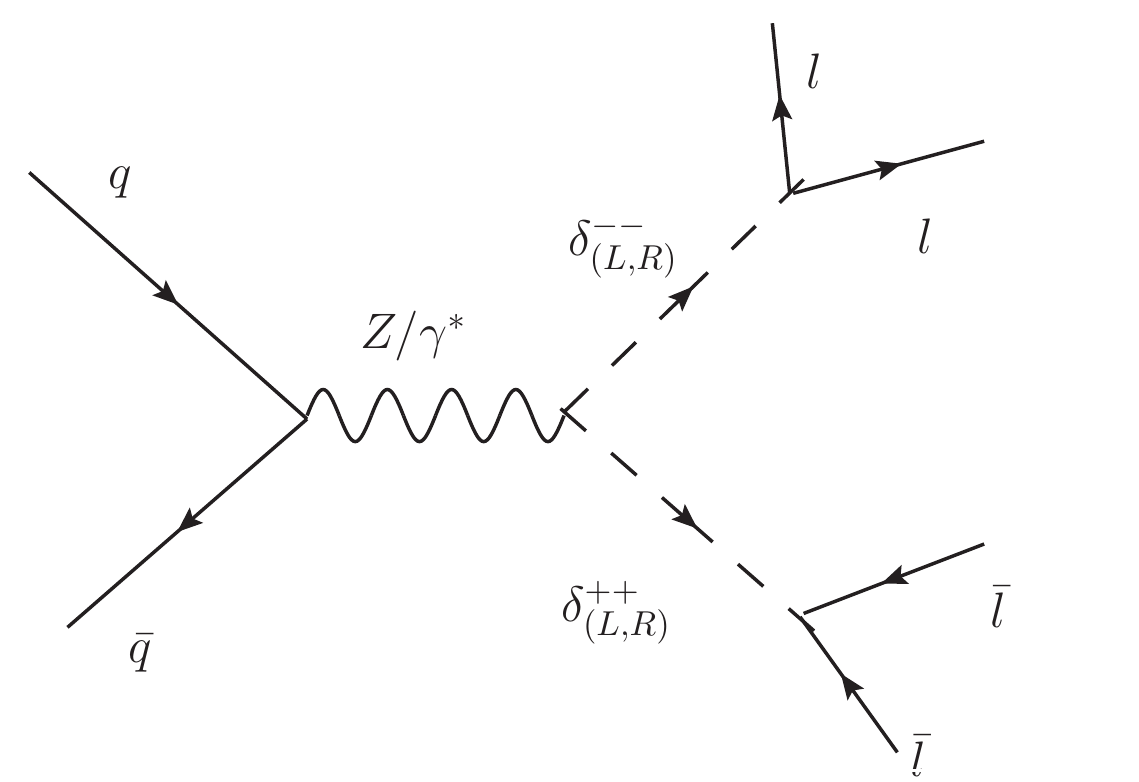} 
\caption{\label{fig2}Pair production of the doubly charged scalars with $Z/\gamma^*$ as  intermediate states.} 
\end{figure}
\begin{figure*}
\includegraphics[width=6in]{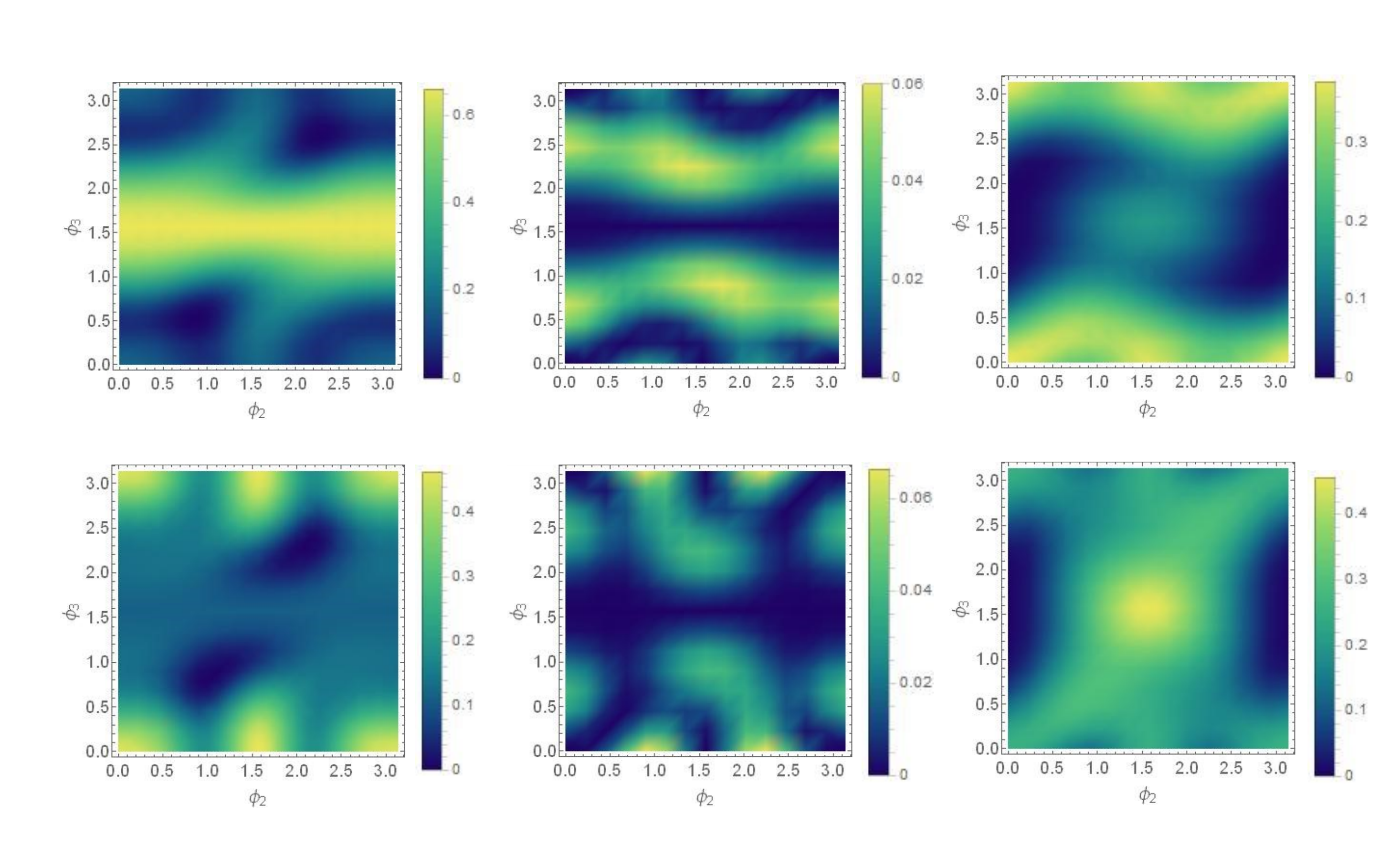} 
\caption{ \label{fig3}\small
Plots for the  branching ratios of $\delta_R^{++}$ into leptons in the $(\phi_2,\phi_3)$ plane. We assume $\delta=\pi/2$ and the masses for the heaviest and lightest right-handed neutrinos, $m_{heaviest}=1$TeV and $m_{lightest}=0.5$TeV in type II dominance. (Left) $\text{Br}(\delta_R^{++} \rightarrow e^+ e^+)$.  
 (Center) $\text{Br}(\delta_R^{++} \rightarrow e^+ \mu^+)$. 
(Right) $\text{Br}(\delta_R^{++} \rightarrow \mu^+ \mu^+)$. (top) Normal hierarchy for neutrino masses. (Bottom)  Inverted hierarchy for neutrino masses.  }
\end{figure*}

In the minimal Left-Right model  the other central role at the LHC is played by the doubly charged scalars \cite{Huitu:1996su,Akeroyd:2005gt,Perez:2008ha,Han:2007bk,Melfo:2011nx}. If light enough they have  interesting signatures at colliders through their  decays into same-sign leptons in the final state. In particular  they  can be produced  with $Z/\gamma^*$ as intermediate states,  see FIG. \ref{fig2}. Pair production has the distinctive signature that consists in same-sign dilepton pairs in the final state. Doubly charged scalars belonging to the $SU(2)_L$ triplet,  should be discovered at the LHC in the lepton-lepton channel.  For $300\text{fb}^{-1}$ of integrated luminosity the mass reach is around 1 TeV.  In the W-W channel  is around   700 GeV \cite{Han:2007bk}. In \cite{Bambhaniya:2014cia}   a the lower bound for $\delta_R^{++}$ of a few hundred GeV (for $v_R \approx 10$TeV) emerge from  the scalar masses assuming $v_1<<v_R$.

The expression for the decay rate of $\delta^{++}_R$ into a lepton pair is

\begin{align}
&\Gamma(\delta_R^{++} \rightarrow l_i^+ l_k^+)= \frac{1}{16\pi(1+\delta_{ik})}|(Y'_{\Delta_R})_{ik}|^2m_{\delta^{++}_R}.  \\& \nonumber \text{(no summation convention over repeated indices)} 
\end{align}

 It can also decay into $W_R^+W_R^+$-pair but this decay is kinetically suppressed if $m_{\delta^{++}_R}<<m_{W_R}$. In this case $\delta^{++}_R$ decays mostly  into leptons and  the branching ratios are

\begin{align}
\frac{\Gamma(\delta_R^{++} \rightarrow l_i^+ l_k^+)}{\Gamma(\delta^{++}_R \rightarrow all)} \equiv &\text{Br}(\delta_R^{++} \rightarrow l_i^+ l_k^+)\nonumber \\  =&\frac{2}{(1+\delta_{ik})}\frac{|(V_R^*m_NV_R^{\dagger})_{ik}|^2}{\sum_{k'} m_{N_{k'}}^2}. \label{brdelta}
\end{align}

 Notice that they  are independent of the $\delta_R^{++}$ mass and  depend in general on the Majorana phases in $K_N$.

Using the parametrization of Eq. \eqref{PMNS} and Eq. \eqref{brdelta}, we compute the branching ratios $\text{Br}(\delta_R^{++} \rightarrow e^+ e^+) $, $\text{Br}(\delta_R^{++} \rightarrow \mu^+ e^+)$ and $\text{Br}(\delta_R^{++} \rightarrow \mu^+ \mu^+)$. In the appendix,   we give the explicit formulas for these branching ratios. In FIG. \ref{fig3}  we plot how the branching ratios depend on the Majorana phases once again assuming type II dominance and $\mathcal{C}$ as the LR symmetry. We do it for the representative values $\delta=\pi/2$, $m_{N_{lightest}}=0.5$TeV and $m_{N_{heaviest}}=1$ TeV,  in both normal and inverted neutrino mass hierarchies.

As we can see from   FIG. \ref{fig3},   the decay rates of $\delta_R^{++}$ into electrons and muons are considerably affected by the Majorana phases $\phi_2$ and $\phi_3$. Notice that  when the branching ratio into two electrons and two muons tends to be large, that of one electron and one muon tends to be smaller. 

  Notice  from Eq. \eqref{brdelta}  that there are  five independent branching ratios into leptons. Taking into account the KS process,  we can see that there are more observables than parameters to be fixed by the experiment (three mixing angles, the Dirac phase $\delta$ and the Majorana phases $\phi_2$ and $\phi_3$). For example, by measuring all the elements of $\hat{V}_R$ through the KS process (as we have explicitly shown) and taking let  say the decays  $\delta_R^{++} \rightarrow e^+ e^+$ and $\delta_R^{++} \rightarrow \mu^+ \mu^+$,  the remaining branching ratios are immediately fixed. This in turn fixes a large number of low-energy experiments,  such as the radiative corrections to muon decay  and the  lepton-flavor-violating decay rates of $\mu \rightarrow e\gamma$, $\mu \rightarrow eee$ and $\mu \rightarrow e$ conversion in nuclei. This is a clear example of  the complementary role played by high and low energy experiment in the determination of the left-right symmetric theory \cite{Tello:2010am,Chakrabortty:2012pp}. 
  
  So far we have considered only the decays of $\delta^{++}_R$ and not $\delta^{++}_L$. The question is whether one can distinguish them without measuring the polarization of the final leptons. We notice that it  can be done at the LHC if $v_L<10^{-4}$ i.e in the leptonic decay region for the doubly charged scalar $\delta_L^{++}$ (see for instance \cite{Perez:2008ha,Melfo:2011nx} for detailed studies on this issue). This is due to the relations  \eqref{relacionCdeltas} and \eqref{relacionPdeltas} and the fact  that the production  cross section  is a factor 2.5 bigger for $\delta_L^{++}$, than the one for $\delta_R^{++}$  \cite{Muhlleitner:2003me,Bambhaniya:2013wza,ATLAS:2014kca}. Of course it is crucial that the backgrounds are negligible after selection criteria are applied \cite{CMS:2012ulp,ATLAS:2014kca}. In \cite{Muhlleitner:2003me},  the next-to-leading order QCD corrections of the production cross-sections at the LHC are calculated and the total theoretical uncertainties are estimated to be $10-15\%$. 
  
 
 At this point the reader may well ask about the physical consequences of the  phases appearing in $K_e$. In this respect we notice that lepton dipole moments and CP-odd asymetries in LFV decays  are in general sensitive to them \cite{Vasquez:2015una}. Then  we can link, in principle, all the parameters appearing in $V_R$ with the experiment.
 
 %
 %

%
%
\section{ Conclusions}
In the context of the minimal Left-Right symmetric theory,  we studied the determination of the leptonic right-handed mixing matrix $V_R$ at the LHC. We considered the Keung-Senjanovi\'c process and the decay of the doubly charged scalar $\delta_R^{++}$.

 For non-degenerate heavy neutrino masses,  the KS process is sensitive to 3 mixing angles and the Dirac-type phase. We proposed a simple approach in order to determine the three mixing angles and the Dirac phase present in $V_R$. This determination may   be done but at least  2 heavy neutrinos must be produced on-shell, in this case the inclusion of tau-leptons in the analysis  is mandatory. For three heavy neutrinos on-shell the three mixing angles and the Dirac phase may be determined by measuring electrons and muons in the final state, rendering the three heavy neutrino case ideal for the LHC. We found  exact analytical solutions for the mixing angles and the Dirac phase $\delta$ in terms of measurable quantities at the LHC in both two and three heavy neutrino cases. We also show that the hadronization effects for the final jets are under control, thus rendering the proposed strategy feasible at the LHC.  Finally we find that for two heavy neutrino at the LHC with  masses near or bellow the TeV,  an integrated luminosity of $224$ $\text{fb}^{-1}$ is required in order to measure the three mixing angles that parametrize the right handed leptonic mixing matrix. In the case of three heavy neutrinos at the LHC and for the range  of heavy neutrino  masses considered (near or bellow the TeV) a luminosity of $68$ $\text{fb}^{-1}$ and $45$ $\text{fb}^{-1}$ is required for both normal and inverted neutrino mass hierarchy respectively. 
 
 For  degenerate heavy neutrinos masses, the lepton-number-violating, same-sign lepton channel (FIG. \ref{fig1}. Bottom)  is in general sensitive to two of the Majorana phases of $V_R$, because in this case there are interference terms between the degenerate 
 right-handed neutrino mass eigenstates.
 
  We point out that the decays of  the doubly charged scalar  $\delta_R^{++}$  into leptons are significantly affected by the same two Majorana phases.  In FIG. \ref{fig3} we show its branching ratios into $e^+e^+$,$e^+\mu^+$ and $\mu^+\mu^+$. We did it for $\mathcal{C}$ as the Left-Right symmetry assuming type 
II see-saw dominance. We considered some representative values of the Dirac phase $\delta$ and the right-handed neutrino masses, in both normal and inverted neutrino mass hierarchies.

 As a consequence of the near equality of the Yukawa couplings of the doubly charged scalars in both parity or charged conjugation as the Left-Right symmetry, the LHC experiment may distinguish $\delta_L^{++}$ from $\delta_R^{++}$ without measuring the polarization of the final-state leptons coming from their decays.

%
%

\section*{Acknowledgements}

The author is deeply grateful to  G. Senjanovi\'c for his teachings, suggestions, advice and encouragement throughout this work. Thanks are also due to S. Bertolini, M. Nemev\v sek and G. Senjanovi\'c for useful and enlightening discussions and  to S. Bertolini, A. Melfo and G. Senjanovi\'c for careful reading of the manuscript.

\begin{widetext}
\section*{Appendix A \label{appendixA}}
In this appendix we present the results for the cross sections obtained from Madgraph 5 \cite{Alwall:2014hca} and Pythia 6 \cite{Sjostrand:2006za}, for different values of the heavy neutrino masses that we used for generation of the relevant processes at the partonic level and the subsequent hadronization effects.

\begin{table}[h]
  \centering
  \begin{tabular}{@{} lccccr @{}}
 \hline
 &  &  Cross section $\sigma$[fb]   \\ 
 & & $m_{N_2} = 1$ TeV \\
  \hspace*{8ex}Processes  & \small$m_{N_1}=100$GeV  & \small$m_{N_1}=500$GeV &  \small$m_{N_1}=750$GeV & \small$m_{N_1}=950$GeV \\ 
 \hline 
    $p p \rightarrow W^+_R\rightarrow N_1 e^+\rightarrow e^+e^+jj$ & 2.53& 2.26  & 2.07 & 1.99 \\ 
     $p p \rightarrow W^+_R\rightarrow N_1 e^+\rightarrow e^+\mu^+jj$& 1.24& 1.11 & 1.01 & 0.98\\ 
     $p p \rightarrow W^+_R\rightarrow N_1 e^+\rightarrow e^+\tau^+jj$ &0.057 & 0.051  & 0.047& 0.045 \\ 
     $p p \rightarrow W^+_R\rightarrow N_1 e^+\rightarrow \mu^+\mu^+jj$ & 0.61 & 0.54  & 0.50 &  0.48\\
     $p p \rightarrow W^+_R\rightarrow N_2 e^+\rightarrow e^+e^+jj$& 0.23 & 0.23 & 0.23 & 0.23 \\
    $p p \rightarrow W^+_R\rightarrow N_2 e^+\rightarrow e^+\mu^+jj$ & 0.29 &  0.29  & 0.29&0.29 \\
     $p p \rightarrow W^+_R\rightarrow N_2 e^+\rightarrow \mu^+\mu^+jj$ & 0.36 &  0.35  & 0.36 & 0.35 \\
 \hline
  \end{tabular}
  \caption{\label{tableI}Cross sections for the different processes considered for two heavy neutrinos at the LHC in the normal hierarchy (NH) neutrino mass spectrum and for different values of the lightest heavy neutrino mass. }
\end{table}

\begin{table}[h]
  \centering
  \begin{tabular}{@{} lccccr @{}}
\hline
 &  &  Cross section $\sigma$[fb]   \\ 
 & & $m_{N_2} = 0.17$ TeV \\
  \hspace*{8ex}Processes  & \small$m_{N_1}=80$GeV  & \small$m_{N_1}=100$GeV &  \small$m_{N_1}=130$GeV & \small$m_{N_1}=160$GeV \\ 
 \hline
    $p p \rightarrow W^+_R\rightarrow N_1 e^+\rightarrow e^+e^+jj$ & 1.64& 1.66  & 1.64 & 1.65 \\ 
     $p p \rightarrow W^+_R\rightarrow N_1 e^+\rightarrow e^+\mu^+jj$& 0.81& 0.82 & 0.80 & 0.81\\ 
     $p p \rightarrow W^+_R\rightarrow N_1 e^+\rightarrow \mu^+\mu^+jj$ &0.40 & 0.40  & 0.39& 0.40 \\ 
     $p p \rightarrow W^+_R\rightarrow N_2 e^+\rightarrow e^+e^+jj$ & 0.21 & 0.21  & 0.21 &  0.21\\
     $p p \rightarrow W^+_R\rightarrow N_2 e^+\rightarrow e^+\mu^+jj$& 0.26 & 0.26 & 0.26 & 0.26 \\
    $p p \rightarrow W^+_R\rightarrow N_2 e^+\rightarrow \mu^+\mu^+jj$ & 0.32 &  0.32  & 0.32 & 0.32 \\
     $p p \rightarrow W^+_R\rightarrow N_3 e^+\rightarrow e^+\mu^+jj$ & 0.29 &  0.15  & 0.16 & 0.17 \\
     $p p \rightarrow W^+_R\rightarrow N_3 e^+\rightarrow \mu^+\mu^+jj$ & 1.02 &  0.51  & 0.55 & 0.58\\
 \hline
  \end{tabular}
  \caption{\label{tableII}Cross sections for the different processes considered for three heavy neutrinos at the LHC in the normal hierarchy (NH) neutrino mass spectrum and for different values of the lightest heavy neutrino mass. }
\end{table}

\begin{table}[h]
  \centering
  \begin{tabular}{@{} lccccr @{}}
\hline
 &  &  Cross section $\sigma$[fb]   
  \\ 
 & & $m_{N_2} = 0.95$ TeV \\
  \hspace*{8ex}Processes   & \small$m_{N_3}=80$GeV  & \small$m_{N_3}=100$GeV &  \small$m_{N_3}=300$GeV & \small$m_{N_3}=500$GeV \\ 
 \hline
    $p p \rightarrow W^+_R\rightarrow N_1 e^+\rightarrow e^+e^+jj$ & 1.97& 2.01  & 1.96 & 1.96 \\ 
     $p p \rightarrow W^+_R\rightarrow N_1 e^+\rightarrow e^+\mu^+jj$&  0.97 & 0.99 & 0.96 & 0.96\\ 
     $p p \rightarrow W^+_R\rightarrow N_1 e^+\rightarrow \mu^+\mu^+jj$ & 0.48 & 0.48  & 0.47& 0.47 \\ 
     $p p \rightarrow W^+_R\rightarrow N_2 e^+\rightarrow e^+e^+jj$ & 0.24 & 0.24  & 0.24 &  0.24\\
     $p p \rightarrow W^+_R\rightarrow N_2 e^+\rightarrow e^+\mu^+jj$& 0.30& 0.30 & 0.30& 0.30 \\
    $p p \rightarrow W^+_R\rightarrow N_2 e^+\rightarrow \mu^+\mu^+jj$ & 0.36 &  0.35  & 0.36 & 0.35 \\
     $p p \rightarrow W^+_R\rightarrow N_3 e^+\rightarrow e^+\mu^+jj$ & 0.24 &  0.24  & 0.22 & 0.23 \\
     $p p \rightarrow W^+_R\rightarrow N_3 e^+\rightarrow \mu^+\mu^+jj$ & 0.84&  0.84  & 0.78 & 0.81\\
 \hline
  \end{tabular}
  \caption{\label{tableIII} Cross sections for the different processes considered for three heavy neutrinos at the LHC in the inverted hierarchy (IH) neutrino mass spectrum and for different values of the lightest heavy neutrino mass. }
\end{table}
\end{widetext}
\newpage
\begin{widetext}
\section*{Appendix B  }

 In this appendix we show the explicit formulas for the branching ratios  $\text{Br}(\delta_R^{++} \rightarrow e^+ e^+) $, $\text{Br}(\delta_R^{++} \rightarrow \mu^+ e^+)$ and $\text{Br}(\delta_R^{++} \rightarrow \mu^+ \mu^+)$,

\begin{align}
\text{Br}(\delta_R^{++} \rightarrow e^+ e^+) =\frac{1}{\sum_k m_{N_k}^2}|c^2_{13}c^2_{12}m_{N_1}+ e^{-2i\phi_2}c_{13}^2s^2_{12}m_{N_2}  + e^{-2i(\phi_3-\delta)}s^2_{13}m^2_{N_3}|^2, \nonumber \\ \label{bree}
\end{align}
\begin{align}
\text{Br}(\delta_R^{++} \rightarrow e^+ \mu^+) =& \frac{2}{\sum_k m_{N_k}^2}|(-s_{12}c_{23}-c_{12}s_{23}s_{13}e^{-i\delta})  c_{12}c_{13}m_{N_1}+ (c_{12}c_{23}-s_{12}s_{23}s_{13}e^{-i\delta})s_{12}c_{13}e^{-2i\phi_2}m_{N_2} \nonumber\\ &+s_{23}c_{13}s_{13}e^{-i(2\phi_3-\delta)}m_{N_3} |^2 , \label{brem}
\end{align}
\begin{align}
 \text{Br}(\delta_R^{++} \rightarrow \mu^+ \mu^+) = & \frac{1}{\sum_k m_{N_k}^2}|(-s_{12}c_{23}-c_{12}s_{23}s_{13}e^{-i\delta})^2m_{N_1} +(c_{12}c_{23}-s_{12} s_{23}s_{13}e^{-i\delta})^2e^{-2i\phi_2}m_{N_2} \nonumber \\ & +s_{23}^2c_{13}^2e^{-2i\phi_3}m_{N_3}|^2. \nonumber \\ \label{brmm}
\end{align}

Notice that this branching ratios are independent of the doubly-charged scalar masses and depend only on the masses of the heavy neutrinos
\end{widetext}
\pagebreak


\begin{thebibliography}{99}
\bibitem{lrmodel}
J.~C.~Pati and A.~Salam,
Phys.\ Rev.\ D {\bf 10} (1974) 275;
R.~N.~Mohapatra and J.~C.~Pati,
  Phys.\ Rev.\  D {\bf 11}, 566 (1975);
R.~N.~Mohapatra and J.~C.~Pati,
  Phys.\ Rev.\  D {\bf 11}, 2558 (1975);
G.~Senjanovi\'c and R.~N.~Mohapatra,
  Phys.\ Rev.\  D {\bf 12}, 1502 (1975).


\bibitem{GoranNuclphys79}
 G.~Senjanovi\'c,
Nucl.\ Phys.\ B {\bf 153} (1979) 334.

 \bibitem{Senjanovic:2011zz} 
 G.~Senjanovi\'c,
 Riv.\ Nuovo Cim.\  {\bf 034}, 1 (2011);
 G.~Senjanovi\'c,
 Int.\ J.\ Mod.\ Phys.\ A {\bf 26}, 1469 (2011)
 [arXiv:1012.4104 [hep-ph]];
V. Tello, PhD Thesis, SISSA (2012)

\bibitem{Minkowski:1977sc}
P.~Minkowski,
Phys.\ Lett.\ B {\bf 67} (1977) 421.

\bibitem{Mohapatra:1979ia}
  R.~N.~Mohapatra and G.~Senjanovi\'c,
  Phys.\ Rev.\ Lett.\  {\bf 44} (1980) 912.
  
\bibitem{Mohapatra:1980yp}
 R.~N.~Mohapatra and G.~Senjanovi\'c,
Phys.\ Rev.\ D {\bf 23} (1981) 165.

\bibitem{see-sawothers1}
S. Glashow, Quarks and leptons, Carg\'ese 1979, ed. M.
L\'evy (Plenum, NY, 1980); M. Gell-Mann et al.,
P. Ramond, R. Slansky, Supergravity Stony Brook work-
shop, New York, 1979, ed. P. Van Niewenhuizen, D. Free-
man (North Holland, Amsterdam, 1980).
T. Yanagida, Workshop on unifed theories and baryon
number in the universe, ed. A. Sawada, A. Sugamoto
(KEK, Tsukuba, 1979).

\bibitem{Beall:1981ze} 
  G.~Beall, M.~Bander and A.~Soni,
  Phys.\ Rev.\ Lett.\  {\bf 48}, 848 (1982). For recent updates see Y.~Zhang, H.~An, X.~Ji and R.~N.~Mohapatra, Nucl.\ Phys.\ B {\bf 802}, 247 (2008).  [arXiv:0712.4218 [hep-ph]] and  A.~Maiezza, M.~Nemev\v sek, F.~Nesti and G.~Senjanovi\'c, Phys.\ Rev.\ D {\bf 82}, 055022 (2010). [arXiv:1005.5160 [hep-ph]].

\bibitem{Bertolini:2014sua} 
S.~Bertolini, A.~Maiezza and F.~Nesti,
Phys.\ Rev.\ D {\bf 89}, 095028 (2014)
[arXiv:1403.7112 [hep-ph]].


  
\bibitem{Maiezza:2014ala} 
  A.~Maiezza and M.~Nemev\v sek,
  arXiv:1407.3678 [hep-ph].
  
  
\bibitem{Khachatryan:2014dka} 
  V.~Khachatryan {\it et al.}  [CMS Collaboration],
  arXiv:1407.3683 [hep-ex].

\bibitem{Keung:1983uu} 
  W.~Y.~Keung and G.~Senjanovi\'c,
  Phys.\ Rev.\ Lett.\  {\bf 50}, 1427 (1983).


  
\bibitem{Aguilar-Saavedra:2014ola} 
  F.~F.~Deppisch, T.~E.~Gonzalo, S.~Patra, N.~Sahu and U.~Sarkar,
  arXiv:1407.5384 [hep-ph];
  J.~A.~Aguilar-Saavedra and F.~R.~Joaquim,
  arXiv:1408.2456 [hep-ph];
  
 \bibitem{Aguilar-Saavedra:2014ola1}
  J.~A.~Aguilar-Saavedra and F.~R.~Joaquim,
  ``A closer look at the possible CMS signal of a new gauge boson,''
  arXiv:1408.2456 [hep-ph];
  
  \bibitem{Heikinheimo:2014tba}
  M.~Heikinheimo, M.~Raidal and C.~Spethmann,
  Eur.\ Phys.\ J.\ C {\bf 74} (2014) 10,  3107
  [arXiv:1407.6908 [hep-ph]].
  
  
\bibitem{Gluza:2015goa} 
  J.~Gluza and T.~Jeliński,
  Phys.\ Lett.\ B {\bf 748}, 125 (2015)
  doi:10.1016/j.physletb.2015.06.077
  [arXiv:1504.05568 [hep-ph]].
  
\bibitem{Dobrescu:2015qna} 
  B.~A.~Dobrescu and Z.~Liu,
  Phys.\ Rev.\ Lett.\  {\bf 115}, no. 21, 211802 (2015)
  doi:10.1103/PhysRevLett.115.211802
  [arXiv:1506.06736 [hep-ph]].
  
\bibitem{Coloma:2015una}
  P.~Coloma, B.~A.~Dobrescu and J.~Lopez-Pavon,
  arXiv:1508.04129 [hep-ph].
  
\bibitem{Bandyopadhyay:2015fka} 
  T.~Bandyopadhyay, B.~Brahmachari and A.~Raychaudhuri,
  arXiv:1509.03232 [hep-ph].
  
\bibitem{Dev:2015pga} 
  P.~S.~Bhupal Dev and R.~N.~Mohapatra,
  Phys.\ Rev.\ Lett.\  {\bf 115}, no. 18, 181803 (2015)
  doi:10.1103/PhysRevLett.115.181803
  [arXiv:1508.02277 [hep-ph]].
  
\bibitem{Brehmer:2015cia} 
  J.~Brehmer, J.~Hewett, J.~Kopp, T.~Rizzo and J.~Tattersall,
  JHEP {\bf 1510}, 182 (2015)
  doi:10.1007/JHEP10(2015)182
  [arXiv:1507.00013 [hep-ph]].
  
  
  

\bibitem{Senjanovic:2014pva} 
  G.~Senjanovi\'c and V.~Tello,
  Phys.\ Rev.\ Lett.\  {\bf 114}, no. 7, 071801 (2015)
  [arXiv:1408.3835 [hep-ph]].
  
\bibitem{Senjanovic:2015yea} 
  G.~Senjanovi\'c and V.~Tello,
  arXiv:1502.05704 [hep-ph].
  
\bibitem{Fowlie:2014mza} 
  A.~Fowlie and L.~Marzola,
  Nucl.\ Phys.\ B {\bf 889}, 36 (2014)
  [arXiv:1408.6699 [hep-ph]].
  
\bibitem{Nemevsek:2012iq} 
  M.~Nemev\v sek, G.~Senjanovi\'c and V.~Tello,
  Phys.\ Rev.\ Lett.\  {\bf 110}, no. 15, 151802 (2013)
  [arXiv:1211.2837 [hep-ph]].
  
 \bibitem{trabajoGVN}
   M.~Nemev\v sek, G.~Senjanovi\'c and V.~Tello. Work to appear.
  
\bibitem{Ferrari:2000sp}
  A.~Ferrari {\it et al.},
  Phys.\ Rev.\  D {\bf 62} (2000) 013001;
  S.~N.~Gninenko, M.~M.~Kirsanov, N.~V.~Krasnikov and V.~A.~Matveev,
  Phys.\ Atom.\ Nucl.\  {\bf 70} (2007) 441;
  M.~Nemev\v sek, F.~Nesti, G.~Senjanovi\'c and Y.~Zhang,
  Phys.\ Rev.\ D {\bf 83}, 115014 (2011)
  [arXiv:1103.1627 [hep-ph]].

\bibitem{Gopalakrishna:2010xm} 
  S.~Gopalakrishna, T.~Han, I.~Lewis, Z.~g.~Si and Y.~F.~Zhou,
  Phys.\ Rev.\ D {\bf 82}, 115020 (2010)
  [arXiv:1008.3508 [hep-ph]].


\bibitem{Han:2012vk} 
  T.~Han, I.~Lewis, R.~Ruiz and Z.~g.~Si,
  Phys.\ Rev.\ D {\bf 87}, 035011 (2013)
  [Erratum-ibid.\ D {\bf 87}, no. 3, 039906 (2013)]
  [arXiv:1211.6447 [hep-ph]].


\bibitem{Vasquez:2015una} 
  J.~C.~Vasquez,
  JHEP {\bf 1509}, 131 (2015)
  [arXiv:1504.05220 [hep-ph]].

\bibitem{Roitgrund:2014zka} 
  A.~Roitgrund, G.~Eilam and S.~Bar-Shalom,
  arXiv:1401.3345 [hep-ph].
  
\bibitem{Alwall:2014hca} 
  J.~Alwall {\it et al.},
  JHEP {\bf 1407}, 079 (2014)
  [arXiv:1405.0301 [hep-ph]].
  
\bibitem{Sjostrand:2006za} 
  T.~Sjostrand, S.~Mrenna and P.~Z.~Skands,
  JHEP {\bf 0605}, 026 (2006)
  [hep-ph/0603175].
  
  
  
\bibitem{Huitu:1996su}
  K.~Huitu, J.~Maalampi, A.~Pietila and M.~Raidal,
  Nucl.\ Phys.\ B {\bf 487} (1997) 27
  [hep-ph/9606311].


\bibitem{Akeroyd:2005gt} 
  A.~G.~Akeroyd and M.~Aoki,
  Phys.\ Rev.\ D {\bf 72}, 035011 (2005)
  [hep-ph/0506176];
  G.~Azuelos, K.~Benslama and J.~Ferland,
  J.\ Phys.\ G {\bf 32}, no. 2, 73 (2006)
  [hep-ph/0503096];
  A.~G.~Akeroyd, M.~Aoki and H.~Sugiyama,
  Phys.\ Rev.\ D {\bf 77}, 075010 (2008)
  [arXiv:0712.4019 [hep-ph]].

\bibitem{Perez:2008ha} 
  P.~Fileviez Perez, T.~Han, G.~y.~Huang, T.~Li and K.~Wang,
  Phys.\ Rev.\ D {\bf 78}, 015018 (2008)
  [arXiv:0805.3536 [hep-ph]].
 
 \bibitem{Han:2007bk} 
  T.~Han, B.~Mukhopadhyaya, Z.~Si and K.~Wang,
  Phys.\ Rev.\ D {\bf 76}, 075013 (2007)
  [arXiv:0706.0441 [hep-ph]].
  
  \bibitem{Melfo:2011nx} 
  A.~Melfo, M.~Nemev\v sek, F.~Nesti, G.~Senjanovic and Y.~Zhang,
  Phys.\ Rev.\ D {\bf 85}, 055018 (2012)
  [arXiv:1108.4416 [hep-ph]].
  
\bibitem{Bambhaniya:2014cia} 
  G.~Bambhaniya, J.~Chakrabortty, J.~Gluza, T.~Jeliński and M.~Kordiaczynska,
  Phys.\ Rev.\ D {\bf 90}, no. 9, 095003 (2014)
  [arXiv:1408.0774 [hep-ph]].
  
  
  
\bibitem{Tello:2010am} 
  V.~Tello, M.~Nemev\v sek, F.~Nesti, G.~Senjanovi\'c and F.~Vissani,
  Phys.\ Rev.\ Lett.\  {\bf 106}, 151801 (2011)
  [arXiv:1011.3522 [hep-ph]];
  M.~Nemev\v sek, F.~Nesti, G.~Senjanovi\'c and V.~Tello,
  arXiv:1112.3061 [hep-ph].
  
\bibitem{Chakrabortty:2012pp} 
  J.~Chakrabortty, J.~Gluza, R.~Sevillano and R.~Szafron,
  JHEP {\bf 1207}, 038 (2012)
  [arXiv:1204.0736 [hep-ph]].
  
\bibitem{Muhlleitner:2003me}
  M.~Muhlleitner and M.~Spira,
  Phys.\ Rev.\ D {\bf 68} (2003) 117701
  [hep-ph/0305288].

\bibitem{Bambhaniya:2013wza} 
  G.~Bambhaniya, J.~Chakrabortty, J.~Gluza, M.~Kordiaczyńska and R.~Szafron,
  JHEP {\bf 1405}, 033 (2014)
  [arXiv:1311.4144 [hep-ph]].
  
\bibitem{ATLAS:2014kca} 
  G.~Aad {\it et al.}  [ATLAS Collaboration],
  JHEP {\bf 1503}, 041 (2015)
  [arXiv:1412.0237 [hep-ex]].
  
\bibitem{CMS:2012ulp}
  CMS Collaboration [CMS Collaboration],
  CMS-PAS-HIG-12-005.
 
\end{thebibliography}
\end{document}